# AN ALTERNATIVE TREATMENT FOR YUKAWA-TYPE POTENTIALS

B. Gönül, K. Köksal and E. Bakır

Department of Engineering Physics, University of Gaziantep, 27310, Gaziantep-Turkey

## Abstract

We propose a new approximation scheme to obtain analytic expressions for the bound state energies and eigenfunctions of Yukawa like potentials. The predicted energies are in excellent agreement with the accurate numerical values reported in the literature.

## 1. INTRODUCTION

Since the pioneering work of Yukawa [1] the potential,

$$V(r) = -(A/r)\exp(-\alpha r) \qquad (1)$$

has been extensively investigated in the literature. This is due to the special role of this potential in different branches of physics. In plasma physics it is known as the Debye-Hückel potential, in solid-state physics and atomic physics it is called the Thomas-Fermi or screened Columb potential. Also, this potential is well known in nuclear physics as the dominant central part of nucleon-nucleon interaction arising out of the one-pion-exchange mechanism. Thus, the parameters $A$ and $\alpha$ are given by different expressions depending on the problem under consideration. In all these cases, a knowledge of the various bound state energies is essential for understanding and correlating the properties of different systems. Since the Schrödinger equation for such potential does not admit exact analytic solutions, various numerical and approximate analytical methods, e.g. [2-6], have been developed in the past.

More recently, a new methodology [7] has been introduced based on the decompose of the radial Schrödinger equation in two pieces having an exactly solvable part with an additional piece leading to either a closed analytical solution or an approximate treatment depending on the nature of the additional/perturbed potential. The applications [8] of this novel treatment to different problems in both, bound and continuum regions, have been proven the success of the formalism. With the confidence gained from these applications, and bearing in mind the significance of a reliable algebraic solution for Yukawa-type potentials, that is clearly reported in [6], we demonstrate here how such interaction potentials can be simply treated within the framework of the present formalism.

The outline of the paper is as follows. In section 2 we sketch the main steps of the procedure. We present only those formulae that are necessary for understanding the following calculations; all details and the mathematical foundations can be found in [7,8]. In section 3 we apply the approach to the Schrödinger equation with the Yukawa potential and present the results obtained analytically, which makes clear that the present scheme gives quite good accuracy for energy values despite its analytical nature. Final section involves some concluding remarks.

## 2. FORMALISM

For the consideration of spherically symmetric potentials, the corresponding Schrödinger equation in the bound state domain for the radial wave function has the form

$$\frac{\hbar^2}{2m}\frac{\psi_n''(r)}{\psi_n(r)} = [V(r) - E_n] \quad , \quad V(r) = \left[V_0(r) + \frac{\hbar^2}{2m}\frac{\ell(\ell+1)}{r^2}\right] + \Delta V(r), \tag{2}$$

where $\Delta V$ is a perturbing potential. Let us write the wave function $\psi_n$ as

$$\psi_n(r) = \chi_n(r)\phi_n(r) \quad , \tag{3}$$

in which $\chi_n$ is the known normalized eigenfunction of the unperturbed Schrödinger equation whereas $\phi_n$ is a moderating function corresponding to the perturbing potential. Substituting (3) into (2) yields

$$\frac{\hbar^2}{2m}\left(\frac{\chi_n''}{\chi_n} + \frac{\phi_n''}{\phi_n} + 2\frac{\chi_n'}{\chi_n}\frac{\phi_n'}{\phi_n}\right) = V - E_n \quad . \tag{4}$$

Instead of setting the functions $\chi_n$ and $\phi_n$, we will set their logarithmic derivatives

$$W_n = -\frac{\hbar}{\sqrt{2m}}\frac{\chi_n'}{\chi_n} \quad , \quad \Delta W_n = -\frac{\hbar}{\sqrt{2m}}\frac{\phi_n'}{\phi_n} \tag{5}$$

which leads to

$$\frac{\hbar^2}{2m}\frac{\chi_n''}{\chi_n} = W_n^2 - \frac{\hbar}{\sqrt{2m}}W_n' = \left[V_0 + \frac{\hbar^2}{2m}\frac{\ell(\ell+1)}{r^2}\right] - \varepsilon_n, \qquad (6)$$

where $\varepsilon_n$ is the eigenvalue of the unperturbed and exactly solvable potential, and

$$\frac{\hbar^2}{2m}\left(\frac{\phi_n''}{\phi_n} + 2\frac{\chi_n'}{\chi_n}\frac{\phi_n'}{\phi_n}\right) = \Delta W_n^2 - \frac{\hbar}{\sqrt{2m}}\Delta W_n' + 2W_n\Delta W_n = \Delta V - \Delta\varepsilon_n, \qquad (7)$$

in which $\Delta\varepsilon_n$ is the eigenvalue for the perturbed potential, and $E_n = \varepsilon_n + \Delta\varepsilon_n$.

If Eq. (7), which is the most significant piece of the present formalism, can be solved analytically having a closed form, as in (6), the whole problem, Eq. (2), simply reduces to

$$(W_n + \Delta W_n)^2 - \frac{\hbar}{\sqrt{2m}}(W_n + \Delta W_n)' = V - E_n, \qquad (8)$$

which is a well known treatment within the frame of supersymmetric quantum theory (SSQT) [9]. Obviously, this famous prescription of the related literature appears naturally as a subset of the present scheme. In addition, it is stressed that Eq. (7) provides a much better framework in treating such problems than the one in the supersymmetric quantum mechanics. Because in SSQT one should start from $W_{n=0}$ corresponding to the ground state wavefunction and use the linear operators $(A^\pm)$ to obtain wavefunctions/superpotentials of the excited states. However, within the frame of present scheme one knows explicitly the solution of Eq. (6), namely the whole spectrum and corresponding eigenfunctions of the unperturbed interaction potential, so that one can easily calculate the required $W_n$ for any state of interest leading to direct computation of related corrections to the unperturbed energy and wavefunction.

For the perturbation technique presented in this letter, we have initially assumed that we could split the given potential in two parts, Eq. (2). The main part corresponds to a shape invariant potential, Eq. (6), for which the superpotential is known analytically and the remaining part is treated as a perturbation, Eq. (7). Therefore, it is obvios that Yukawa like potentials can be treated using our prescription. In this case, the zeroth-order term corresponds to the Coulomb potential while higher-order terms consitute the perturbation. However, the perturbation term in its present form cannot be solved exactly through Eq. (7). Thus, one should expand the functions related to the perturbation in terms of the perturbation parameter $\lambda$,

$$\Delta V(r;\lambda) = \sum_{k=1}^{\infty}\lambda^k \Delta V_k(r), \quad \Delta W_n(r;\lambda) = \sum_{k=1}^{\infty}\lambda^k \Delta W_{nk}(r), \quad \Delta\varepsilon_n(\lambda) = \sum_{k=1}^{\infty}\lambda^k \Delta\varepsilon_{nk} \qquad (9)$$

where $k$ denotes the perturbation order. Substitution of the above expansion into Eq. (7) by equating terms with the same power of $\lambda$ on both sides yields up to $O(\lambda^3)$

$$2W_n \Delta W_{n1} - \frac{\hbar}{\sqrt{2m}} \Delta W'_{n1} = \Delta V_1 - \Delta \varepsilon_{n1} \quad , \tag{10}$$

$$\Delta W_{n1}^2 + 2W_n \Delta W_{n2} - \frac{\hbar}{\sqrt{2m}} \Delta W'_{n2} = \Delta V_2 - \Delta \varepsilon_{n2} \quad , \tag{11}$$

$$2(W_n \Delta W_{n3} + \Delta W_{n1} \Delta W_{n2}) - \frac{\hbar}{\sqrt{2m}} \Delta W'_{n3} = \Delta V_3 - \Delta \varepsilon_{n3} \quad . \tag{12}$$

Eq. (7) and its expansion, Eqs. (10-12), give a flexibility for the easy calculations of the perturbative corrections to energy and wave functions for the *nth* state of interest through an appropriately chosen perturbed superpotential, unlike the other perturbation theories. We will show in the next section through the applications that this feature of the present model leads to a simple framework in obtaining the corrections to all states without using complicated and tedious mathematical procedures.

## 3. APPLICATION

We now apply this method to a Yukawa-type potential with the angular momentum barrier

$$V = -\left(\frac{A}{r}\right)\exp(-\alpha r) + \frac{\ell(\ell+1)\hbar^2}{2mr^2} = \left[V_0 + \frac{\ell(\ell+1)\hbar^2}{2mr^2}\right] + \Delta V \quad , \tag{13}$$

where the first piece is the zeroth-order and shape invariant exactly solvable piece corresponding to the unperturbed potential with $V_0 = -A/r$ while $\Delta V$ is the perturbation term $\Delta V = A\alpha - (A\alpha^2/2)r + (A\alpha^3/6)r^2 - (A\alpha^4/24)r^3 + \cdots$ through the expansion of the exponential term. Our careful calculations have clarified that the main contributions come from the first three terms. Hence, the present calculations are performed up to the second-order involving only these additional potential terms, which suprisingly provide highly accurate results.

### 3.1. Ground state calculations $(n = 0)$

In the light of Eq. (6), the zeroth-order calculations leading to exact solutions can be carried out readily with the choice of a suitable $W_{n=0}$ yielding the Coulomb potential,

$$W_{n=0}(r) = -\frac{\hbar}{\sqrt{2m}}\frac{\ell+1}{r} + \sqrt{\frac{m}{2}}\frac{A}{(\ell+1)\hbar} \quad , \quad \varepsilon_n = -\frac{mA^2}{2\hbar^2(n+\ell+1)^2} \quad , \quad n = 0,1,2,\ldots$$

$$\chi_n(r) = \left[\frac{2mA}{(n+\ell+1)\hbar^2}\right]^{\ell+1} \left(\frac{1}{n+\ell+1}\right) \frac{r^{\ell+1}}{\sqrt{\frac{\hbar}{mAn!}(n+2\ell+1)!}} \exp\left[-\frac{mA}{(n+\ell+1)\hbar^2}r\right] \times L_n^{2\ell+1}\left[\frac{2mA}{(n+\ell+1)\hbar^2}r\right]$$

(14)

in which

$$L_n^\ell(r) = \sum_{m=0}^{n} \frac{\Gamma(n+\ell+1)(-r)^m}{\Gamma(m+\ell+1)(n-m)!m!}$$ is an associate Laguarre polynomial. These analytical solutions are already exist in the literature, providing a superiority to the present calculations.

For the calculation of corrections to the zeroth-order energy and wavefunction, one needs to consider the expressions leading to the first- and second-order perturbation given by Eqs. (10-11). Multiplication of each term by $\chi_n^2$ in these equations, and keeping in mind the relation $W_n = -\frac{\hbar}{\sqrt{2m}}\frac{\chi_n'}{\chi_n}$ in Eq. (5), one can obtain general expressions for the corrections in the first-

$$\Delta\varepsilon_{n1} = \int_{-\infty}^{\infty} \chi_n^2(r)\left(-\frac{A\alpha^2}{2}r\right)dr \quad , \quad \Delta W_{n1}(r) = \frac{\sqrt{2m}}{\hbar}\frac{1}{\chi_n^2(r)}\int^r \chi_n^2(z)\left(\Delta\varepsilon_{n1} + \frac{A\alpha^2}{2}z\right)dz \quad , \quad (15)$$

and second-order

$$\Delta\varepsilon_{n2} = \int_{-\infty}^{\infty} \chi_n^2(r)\left[\frac{A\alpha^3}{6}r^2 - \Delta W_{n1}^2(r)\right]dr \quad ,$$

$$\Delta W_{n2}(r) = \frac{\sqrt{2m}}{\hbar}\frac{1}{\chi_n^2(r)}\int^r \chi_n^2(z)\left[\Delta\varepsilon_{n2} + \Delta W_{n1}^2(z) - \frac{A\alpha^3}{6}z^2\right]dz \quad , \quad (16)$$

for any state of interest. According to these formulas, we can calculate $\Delta W_{n1}$ and $\Delta W_{n2}$ explicitly only when we know what the energy corrections $\Delta\varepsilon_{n1}$ and $\Delta\varepsilon_{n2}$ are, from which the whole of the perturbed wavefunction can be calculated in a closed form by Eq. (5). It is also noted that the lower limit of the integration for energy calculations should be changed from $-\infty$ to $0$ to accomodate the fact that $r$ is always positive.

Thus, the ground state calculations within the frame of Eqs. (15) and (16) give

$$\Delta\varepsilon_{01} = -\frac{\hbar^2(\ell+1)(2\ell+3)}{4m}\alpha^2 \quad , \quad \Delta\varepsilon_{02} = \frac{\hbar^4(\ell+1)^2(\ell+2)(2\ell+3)}{12Am^2}\alpha^3 - \frac{\hbar^6(\ell+1)^4(\ell+2)(2\ell+3)}{16A^2m^3}\alpha^4$$

$$\Delta W_{01}(r) = -\frac{(\ell+1)\hbar\alpha^2}{2\sqrt{2m}}r \quad , \quad \Delta W_{02}(r) = \frac{\{(\ell+1)\hbar\alpha^3 r[mAr + (\ell+1)(\ell+2)\hbar^2[3\alpha\hbar^2(\ell+1)^2 - 4mA]]\}}{24\sqrt{2m}(mA)^2}$$

(17)

The analytical expressions for the lowest energy and radial wavefunction of a Yukawa-type potential are then given by

$$E_{n=0} \approx \varepsilon_{n=0} + A\alpha + \Delta\varepsilon_{01} + \Delta\varepsilon_{02}, \quad \psi_{n=0}(r) \approx \chi_n(r)\phi_n(r), \quad \phi_{n=0}(r) \approx \exp\left(-\frac{\sqrt{2m}}{\hbar}\int^r (\Delta W_{01} + \Delta W_{02})dz\right)$$
(18)

These explicit expressions support the similar works in [4-6]. Table 1 shows numerical values of the perturbed energies for a few values of $n$ and $\alpha$. The results obtained are compared with those of [4,5], together with the results of Rogers and his co-workers [2] who solved the Schrödinger equation numerically. Our results are in remarkably good agreement. Table 2 and 3 illustrate another comparison of our calculations with those of [6] who carried out their calculations in a different unit. These two different comparison make clear the sensitivity of present calculations although the procedures in [4-6] to reproduce the corrections analytically seem similar to ours. In particular, Table 2 clarifies that the present method is a very useful one for large potential parameters ($A$), for which numerical solution of the Schrödinger equation is extremely difficult. Because, for a large strength the Yukawa potential is very deep and the wavefunction becomes very sharply peaked near the origin. This causes a great deal of difficulty in the numerical solution of the Schrödinger equation, which is reflected in the instability of the wavefunction thus obtained, although the energy eigenvalue is fairly stable and accurate.

## 3.2. Excited state calculations $(n \geq 1)$

The procedure leads to a handy recursion relations in the case of ground states, but becomes extremely cumbersome in the description of radial excitations when nodes of wavefunctions are taken into account, in particular during the higher order calculations. Although several attempts have been made to by pass this difficulty and improve calculations in dealing with excited states, e.g. [5] within the frame of supersymmetric quantum mechanics and, the related references therein regarding the Logarithmic Perturbation Theory, they have not resulted in a desirable simple algorithm. Hence, as an another objective of this paper, here an explicit treatment is introduced within the frame of the present formalism and described a straightforward procedure for obtaining the perturbation corrections through handy recursion formulae, having the same form both for ground and excited states.

Using Eqs. (5) and (14), the function $W_n$ related to the excited states can be calculated explicitly for the compution of perturbations expressed by (15) and (16). So the first-order corrections in the first excited state $(n=1)$ are

$$\Delta\varepsilon_{11} = -\frac{\hbar^2(\ell+4)(2\ell+3)}{4m}\alpha^2,$$

$$\Delta W_{11}(r) = -\frac{(\ell+2)\hbar\alpha^2 r\left[A^2m^2r^2 - (\ell+1)(2\ell+5)\hbar^2 mAr + (\ell+1)^2(\ell+2)(\ell+4)\hbar^4\right]}{2\sqrt{2m}\left[Amr - (\ell+1)(\ell+2)\hbar^2\right]^2} \quad . \tag{19}$$

However, higher-order calculations have singularity problems during the integrations because of the nodes appearing in $\Delta W_{11}$. To overcome this problem, we focus on a hidden relation in the above equation and with some confidence suggest that

$$\left[Amr - (\ell+1)(\ell+2)\hbar^2\right]^2 \approx \left[A^2m^2r^2 - (\ell+1)(2\ell+5)\hbar^2 mAr + (\ell+1)^2(\ell+2)(\ell+4)\hbar^4\right], \tag{20}$$

which transforms Eq. (19) into

$$\Delta W_{11}(r) \approx -\frac{(\ell+2)\hbar\alpha^2}{2\sqrt{2m}}r \quad . \tag{21}$$

Use of the approximated $\Delta W_{11}$ in (16) gives the energy correction in the second-order as

$$\Delta\varepsilon_{12} = \frac{\hbar^4(\ell+2)^2(\ell+7)(2\ell+3)}{12Am^2}\alpha^3 - \frac{\hbar^6(\ell+2)^4(\ell+7)(2\ell+3)}{16A^2m^3}\alpha^4 \quad . \tag{22}$$

Therefore, the approximate energy value of the Yukawa potential corresponding to the first excited state is

$$E_{n=1} \approx \varepsilon_{n=1} + A\alpha + \Delta\varepsilon_{11} + \Delta\varepsilon_{12} \quad . \tag{23}$$

The related radial wavefunction can be expressed in an analytical form in the light of Eqs (15), (16) and (18), if required.

One can easily check the similarities, as in (20), between the terms of any $W_n = -\frac{\hbar}{\sqrt{2m}}\frac{\chi'_n}{\chi_n}$ of interest to by pass the nodal difficulty as in the first excited state. Our careful and exhausted investigations have revealed that the ratio between these similar terms in $W_n$ for any state is approximately 1, which means that the appromation used here would not affect considerably the sensitivity of the calculations. Furthermore, these investigations put forward an interesting hiearachy between $\Delta W_{n1}$ terms of different quantum states in the first order, circumventing the nodal difficulties elegantly,

$$\Delta W_{nl}(r) \approx -\frac{(n+\ell+1)\hbar\alpha^2}{2\sqrt{2m}} r \quad , \tag{24}$$

which, for instance, for the second excited state $(n=2)$ leads to

$$\Delta\varepsilon_{21} = -\frac{\hbar^2 (2\ell^2 + 17\ell + 27)}{4m}\alpha^2,$$

$$\Delta\varepsilon_{22} = \frac{\hbar^4 (\ell+2)(\ell+3)^2 (2\ell+23)}{12Am^2}\alpha^3 - \frac{\hbar^6 (\ell+2)(\ell+3)^4 (2\ell+23)}{16A^2 m^3}\alpha^4 \quad . \tag{24}$$

Calculations for higher excited states can be carried out in the same manner without employing tedios integrals, results of which are fairly in good agreement with the accurate numerical integration results, see Tables (1) and (3), when compared to the other theoretical works. Finally, though the comparison of our results with those of [2] for large $n-$ and $\ell-$values yields indeed excellent results, we do not illusrate these tables here for clarity, which may be reproduced easily within the scheme described in this section.

## 4. CONCLUDING REMARKS

In conclusion, a new useful technique for solving the bound state problem for Yukawa-type potentials within the frame of Riccati equation have been obtained and the comparison of calculation results with the accurate numerical values has been proven the success of the formalism. Avoiding the disadvantages of the standard non-relativistic perturbation theories, the present formulae have the same simple form both for ground and excited states and provide, in principle, the calculation of the perturbation corrections up to an arbitrary order in analytical or numerical form.

Additionally, the application of the present technique to Yukawa potential is really of great interest leading to analytic expressions for both energy values and wavefunctions. Of particular importance is the apperance of ground state energy in a simple form. Comparing various energy levels with different works in the literature we feel that our analytic treatment quite reliable and further analytic calculation with this non-perturbative scheme would be useful. In particular, our method becomes more reliable as the potential strength increases while the numerical solution of the Schrödinger equation gets unstable and unreliable in calculating especially the wavefunction. Thus, the present method nicely complements the existing numerical methods.

Finally, it is noted that the present results can be extended easily to $N-$dimension with the consideration of the work in [10] by the replacement of the angular momentum term $\ell$ with $\Lambda = (M-3)/2$ where $M = N + 2\ell$.

# TABLES

**Table 1.** Energy eigenvalues of the Yukawa potential in units of $\hbar = m = 1$. For comparison, we set $A = \sqrt{2}$ and $\alpha = gA$.

| State | g | Present Calculations | Ref. [2] (Exact) | Ref. [4] | Ref. [5] |
|---|---|---|---|---|---|
| 1s | 0.002 | -0.99601 | -0.9960 | -0.99601 | -0.9960 |
|  | 0.005 | -0.99004 | -0.9900 | -0.99004 | - |
|  | 0.01 | -0.98015 | -0.9801 | -0.98015 | -0.9801 |
|  | 0.02 | -0.96059 | -0.9606 | -0.96059 | -0.9606 |
|  | 0.025 | -0.95092 | -0.9509 | -0.95092 | - |
|  | 0.05 | -0.90363 | -0.9036 | -0.90363 | -0.9036 |
| 2s | 0.002 | -0.24602 | -0.2460 | -0.24602 | -0.2460 |
|  | 0.005 | -0.24015 | -0.2401 | -0.24015 | - |
|  | 0.01 | -0.23059 | -0.2306 | -0.23058 | -0.2306 |
|  | 0.02 | -0.21230 | -0.21230 | -0.21229 | -0.2124 |
|  | 0.025 | -0.20355 | -0.2036 | -0.20352 | - |
|  | 0.05 | -0.16351 | -0.1635 | -0.16325 | -0.1650 |
| 2p | 0.002 | -0.24602 | -0.2460 | -0.24602 | -0.2460 |
|  | 0.005 | -0.24012 | -0.2401 | -0.24012 | - |
|  | 0.01 | -0.23049 | -0.2305 | -0.23049 | -0.2305 |
|  | 0.02 | -0.21192 | -0.2119 | -0.21193 | -0.2120 |
|  | 0.025 | -0.20299 | -0.2030 | -0.20299 | - |
|  | 0.05 | -0.16144 | -0.1615 | -0.16155 | -0.1625 |
| 3p | 0.002 | -0.10716 | -0.1072 | -0.10716 | -0.1072 |
|  | 0.005 | -0.10142 | -0.1014 | -0.10142 | - |
|  | 0.01 | -0.09231 | -0.09231 | -0.09236 | -0.09236 |
|  | 0.02 | -0.07570 | -0.07570 | -0.07563 | -0.07611 |
|  | 0.025 | -0.06814 | -0.06816 | -0.06799 | - |
|  | 0.05 | -0.03739 | -0.03712 | -0.03486 | -0.04236 |
| 3d | 0.002 | -0.10715 | -0.1072 | -0.10715 | -0.1072 |
|  | 0.005 | -0.1014 | -0.1014 | -0.10137 | - |
|  | 0.01 | -0.09212 | -0.09212 | -0.09212 | -0.09216 |
|  | 0.02 | -0.07502 | -0.07503 | -0.07504 | -0.07531 |
|  | 0.025 | -0.06713 | -0.06715 | -0.06718 | - |
|  | 0.05 | -0.03388 | -0.03383 | -0.03477 | -0.03736 |

**Table 2.** The same as in Table 1, but $\hbar = 2m = 1$, $\alpha = 0.2$ $fm^{-1}$ and $n = 0$.

| A | $\ell$ | Present Calculations | Ref. [6] (Numerical) | Ref. [6] (Analytical) |
|---|---|---|---|---|
| 4 | 0 | -3.2563 | -3.2565 | -3.2199 |
| 8 | 0 | -14.4581 | -14.4571 | -14.4199 |
|   | 1 | -2.5830 | -2.5836 | -2.4332 |
| 16 | 0 | -60.8590 | -60.8590 | -60.8193 |
|   | 1 | -12.9908 | -12.9910 | -12.8375 |
| 24 | 0 | -139.2590 | -139.2594 | -139.2201 |
|   | 1 | -31.3937 | -31.3938 | -31.2385 |
|   | 2 | -11.5951 | -11.5959 | -11.2456 |

**Table 3.** The same as in Table 2, but $n \succ 0$

| A | $\ell$ | n | Present Calculations | Ref. [6] (Numerical) | Ref. [6] (Analytical) |
|---|---|---|---|---|---|
| 16 | 0 | 1 | -13.0271 | -13.0273 | -13.0326 |
|   | 0 | 2 | -4.3937 | -4.3720 | -4.4057 |
|   | 1 | 1 | -4.3612 | -4.3480 | -4.3886 |
| 24 | 0 | 1 | -31.4311 | -31.4313 | -31.4356 |
|   | 0 | 2 | -11.6992 | -11.6998 | -11.7093 |
|   | 0 | 3 | -5.0448 | -5.0442 | -5.0590 |
|   | 0 | 4 | -2.2194 | -2.2033 | -2.2237 |
|   | 1 | 1 | -11.6645 | -11.6653 | -11.6839 |
|   | 1 | 2 | -5.0133 | -5.0135 | -5.0541 |
|   | 1 | 3 | -2.1908 | -2.1770 | -2.2414 |
|   | 2 | 1 | -4.9504 | -4.9516 | -5.0085 |
|   | 2 | 2 | -2.1337 | -2.1241 | -2.2428 |